# A model of $\bar{n}$ annihilation in experimental searches for $n \to \bar{n}$ transformations

E. S. Golubeva[1], J. L. Barrow[2], C. G. Ladd[2]

[1]*Institute for Nuclear Research, Russian Academy of Sciences, Prospekt 60-letiya Oktyabrya 7a, Moscow, 117312, Russia*

[2]*University of Tennessee, Department of Physics, 401 Nielsen Physics Building, 1408 Circle Drive, Knoxville, TN 37996, USA*

Searches for baryon number violation, including searches for proton decay and neutron-antineutron transformation ($n \to \bar{n}$), are expected to play an important role in the evolution of our understanding of beyond Standard Model physics. The $n \to \bar{n}$ is a key prediction of certain popular theories of baryogenesis, and the experiments such as the Deep Underground Neutrino Experiment and the European Spallation Source plan to search for this process with bound- and free-neutron systems. Accurate simulation of this process in Monte Carlo will be important for the proper reconstruction and separation of these rare events from background. This article presents developments towards accurate simulation of the annihilation process for use in a cold, free neutron beam for $n \to \bar{n}$ searches from $\bar{n}C$ annihilation, as $^{12}_{6}C$ is the target of choice for the European Spallation Source's NNBar Collaboration. Initial efforts are also made in this paper to perform analogous studies for intra-nuclear transformation searches in $^{40}_{18}Ar$ nuclei.

## I. INTRODUCTION

### A. Background

As early as 1967, A. D. Sakharov pointed out [1] that for the explanation of the Baryon Asymmetry of the Universe (BAU) there should exist interactions in which baryonic charge is violated besides mere departures from thermal equilibrium and $CP$ symmetry. Thus, experimental searches for baryon number ($B$) violating processes, and in particular the baryon minus lepton ($B - L$) number violating process of neutron—antineutron oscillation ($n \to \bar{n}$), are of great importance due to their possible connections to the explanations of the observed matter-antimatter asymmetry of the universe—as first laid out by V. A. Kuzmin [2] and followed in developments by many authors, see e.g. recent reviews [3-5].

Thus, the search for $n \to \bar{n}$, along with nucleon decay, remains one of the most important areas of modern physics, hopefully leading to an understanding of phenomena related to the BAU.

The best lower limit on a measurement of the oscillation period with free neutrons, $\tau_{n \to \bar{n}}$, was attained at a reactor at the Institut Laue-Langevin (ILL) [6] in Grenoble, France, with a cold neutron beam. These neutrons flew through an evacuated, magnetically shielded pipe of $76\,m$ in length (corresponding to a flight time of $\sim 0.1\,s$), until being allowed to hit a target of carbon ($^{12}_{6}C$) foil (with a thickness of $\sim 130\,\mu m$). This foil would have absorbed antineutrons, resulting in matter-antimatter annihilation which was expected to yield a signal with a star-like topology made of several pions. Particle detectors and calorimeters surrounded the target to record such annihilation events, and was capable of reconstructing the vertex of the pion-star within the central plane of the $^{12}_{6}C$ foil along with the visible energy. In total, the target received $\sim 3 \times 10^{18}$ neutrons, with no recorded annihilation events, i.e. with zero background. This was due to an analysis scheme requiring two or more tracks ($\bar{n}$-annihilation or background-produced mesons, or their decay products) to be reconstructed in the detector as emanating from the $^{12}_{6}C$ foil. As a result, the oscillation limit for free neutrons was established to be

$$\tau_{n \to \bar{n}} \geq 0.86 \times 10^{8}\,s. \qquad (1)$$

In the last two decades since obtaining this result, there have been significant technological developments within the field which have permitted the planning of another transformation experiment, recently proposed at the currently under construction European Spallation Source (ESS) [5,7,8]. According to preliminary estimates, such an experiment could explore this process with $2 - 3$ orders of magnitude higher sensitivity than in [6], leading next generation free neutron experiments to be sensitive to oscillation time range $\tau_{n \to \bar{n}} \sim 10^9 - 10^{10} s$.

Another way to detect $n \to \bar{n}$ is through intra-nuclear searches, and discovery is tantalizing possible. Searches for $\tau_{n \to \bar{n}}$ can be performed in experiments with large underground detectors looking for any hints of the instability of matter. Within the nucleus, spontaneous $\bar{n}$ production would lead to annihilation with another neighboring nucleon, resulting in the release of $\sim 2 \ GeV$ of total energy. However, such intra-nuclear transformations are significantly suppressed compared to $n \to \bar{n}$ in vacuum [5,9-13]. The limit on the $n \to \bar{n}$ intra-nuclear transformation time (in *matter*) $\tau_m$ is associated with the square of the free transformation time [5] through a dimensional suppression factor, $R$:

$$\tau_m = R \cdot \tau_{n \to \bar{n}}^2 \qquad (2)$$

In the nucleus, this suppression is due to differences between the neutron and antineutron nuclear potentials; however, in high mass detectors, this suppression can be compensated by the large number of neutrons available for investigation within the large detector volume. A number of nucleon decay search collaborations have been involved in the search for $n \to \bar{n}$ in nuclei, such as Frejus [14] and Soudan-2 [15] in $^{56}_{26}Fe$, and IMB [16], Kamiokande [17], and Super-Kamiokande (SK) [18] in $^{16}_{8}O$; there has also been a deuteron search performed at SNO [19]. In the Soudan-2 experiment, there is a limit on the transformation time in iron nuclei of $\tau_{Fe} \geq 7.2 \times 10^{31} \ yrs$ [15], which is in line with the limit for the free transformation time of $\tau_{n \to \bar{n}} \geq 1.3 \times 10^8 s$. In SK, which *extracted* 24 $n \to \bar{n}$ candidate events while *expecting* a background count of 24.1 atmospheric neutrino events, these limits were $\tau_O \geq 1.9 \times 10^{32} \ yrs$ [18] and $\tau_{n \to \bar{n}} \geq 2.7 \times 10^8 s$, respectively.

The prevalence of background within SK and other large underground detectors, possibly shrouding a true event, prioritizes the rigorous modeling of both signal and background within an intra-nuclear context. Without any significant improvement in the separation of signal to background in new experiments, it will be possible to *improve* the appearance limit, but impossible to claim any *real discovery*. This contrasts the tantalizing figure that future experiments in large underground detectors could improve the restrictions on processes where $\Delta B = \pm 2$ up to $\sim 10^{33} - 10^{35} \ yrs$ [5] in the absence of background. An experiment possibly capable of such a search for $n \to \bar{n}$ within the $^{40}_{18}Ar$ nucleus is currently under construction, using large liquid argon ($^{40}_{18}Ar$) time projection chamber: the Deep Underground Neutrino Experiment (DUNE) [20].

Whether or not $n \to \bar{n}$ is definitively observed above background in intra-nuclear experiments depends critically upon the separability of signal from background and the energy scale at which the new BSM mechanism will appear. In the case of an observation in intra-nuclear experiments the results will be of great importance for the understanding of fundamental properties of matter, along with building a precise theoretical model describing these properties. Although in the free neutron search [6] no background was detected, the question of background separation might become essential with the planned increase in sensitivity in searches using both free neutrons produced by spallation and bound neutrons in underground experiments, meriting further study beyond this work.

Thus, one requires detailed information about the processes during the annihilation of slow antineutrons on nuclei. The purpose of this work is to create a model describing the annihilation of a slow antineutron incident upon a $^{12}_{6}C$ nucleus for the upcoming transformation experiment using a free neutron beam at ESS. Also, the first steps have also been taken towards a full, realistic simulation of the annihilation resulting from $n \to \bar{n}$ within $^{40}_{18}Ar$ nuclei for DUNE.

### B. Past simulation for free and bound $n \to \bar{n}$ searches

In general, the experiment requires maximum efficiency for detection and reconstruction of incredibly rare antineutrons to be separated from background. The development of Monte Carlo (MC) generators for $n \rightarrow \bar{n}$ searches is not new, and has been an integral part of all past experiments. Sadly, the descriptions of these MCs, as known, are not always complete or seemingly consistent, and are not easily accessible. Information about the generator developed for the ILL experiment [6] is few and far between, unavailable [21], and lacking [22] in detailed explanation.

Intra-nuclear searches have been completed far more times than free neutron experiments, and so their accompanying generators are similarly abundant. Never-the-less, many of their descriptions are scattered throughout a multitude of dissertations and are poorly defended within published works. Similarly, open access to these simulations is lacking. For instance, SK [18] cites only three works in reference to their generator, one of which is a previous work of this paper's lead author, and two of which contain rather ancient antiproton annihilation data; how *exactly* these are implemented within their model is not available.

The authors are also aware of Hewes' work in relation to $n \rightarrow \bar{n}$ in DUNE [23]. However, there exist similar issues to those seen in [18], among them the assumption that the annihilation occurs along the density distribution of the nucleus despite the supposed use of work in [13]. In both [18] and [23], only ∼10 of exclusive annihilation channels are used, whereas our model utilizes ∼100 derived from experimental and theoretical techniques. No previous studies are known to have been tested on their ability to reproduce antinucleon annihilation data, which is a central feature of our work. Our present model is also the first published to incorporate a proper description of the annihilation's dependence on the interaction radius within carbon. Work is underway on a proper implementation of this concept within $^{40}_{18}Ar$.

## C. This work and its goals

Our goal is to create an adequately accurate generator, one which can serve as a platform to be used within all free and intra-nuclear $n \rightarrow \bar{n}$ experiments. In this article, we present the main framework and approaches underlying the model, wherein the annihilation of an antineutron on the target nucleus is considered to consist of several sequential and independent stages. We use the approach originally undertaken in [24,25].

In the first stage of this approach, one defines the absorption point of an antineutron by the nucleus in the framework of the optical model. Our modeling was performed for $10\ meV$ antineutrons incident upon a $^{12}_6C$ nucleus [24,25]. For $^{40}_{18}Ar$, $n \rightarrow \bar{n}$ is assumed to occur within the nucleus, where the nucleons have some Fermi motion, and the present paper shows some first steps in this direction; the process of $n \rightarrow \bar{n}$ within $^{40}_{18}Ar$ will be the focus of our future work. After the point of these quite different initial conditions, all of the following stages of the process for both $^{12}_6C$ and $^{40}_{18}Ar$ do not differ and are considered within a unified approach.

The second stage in this approach is the actual annihilation of the antineutron with one of the constituent intra-nuclear nucleons. In contrast to [24,25], where a statistical model for nucleon-antinucleon annihilation into pions was used, the present paper instead uses a *combined* approach first proposed in [26] and will be described in Section II D. In this paper we use a version of the annihilation model originating in 1992, utilizing corresponding experimental data available at that time. While there do exist more recent findings from later analyses of LEAR data, these are not many in number and will not greatly affect the conclusions reached for ESS from the model, as these can only slightly modify the probabilities of various annihilation channels within the database of our simulation; these can be updated at a later time when we seek even greater precision for $^{40}_{18}Ar$.The third stage is the intra-nuclear cascade (INC), initiated by the emergence and nuclear transport of mesons from the annihilation; decays of short-lived resonances are also handled. In this paper, we use the original version of the model which takes into account the nonlinear effect of decreasing the nuclear density, along with a time coordinate [27], which is necessary for the correct description of the passage of resonances through the nucleus.

The final stage is the de-excitation of the residual nucleus.

In this paper, we present a general description of the model and the first results obtained for $\bar{n}C$ in preparation for the forthcoming ESS experiment. In future developments, the description of individual stages of the process can be modified and improved, but the approach remains the same.

The outline of this paper is as follows: Section II provides a detailed description of the model for all successive stages of the process under consideration. In Section III, a comparison will be made between simulation and experiment to test the model against existing at-rest $\bar{p}C$ annihilation data. In Section IV, some validation tests of the $\bar{n}C$ annihilation event generator output data are shown. In Section V, we summarize our work, and briefly consider a future path toward simulation of intra-nuclear transformations in $^{40}_{18}Ar$.

## II. THE MODEL OF THE ANTINEUTRON ANNIHILATION ON THE NUCLEUS
### A. Absorption of the slow antineutron by the $^{12}_{6}C$ nucleus

In this work, we simulate the annihilation of a cold $(\sim 10\,meV)$ $\bar{n}$ on a $^{12}_{6}C$ nucleus. The calculation of the total annihilation cross section of an $\bar{n}$ on $^{12}_{6}C$ is a separate problem that is not considered within the scope of this model,, and instead the annihilation event itself is the starting point.

The interaction of a slow antineutron with the nucleus cannot be considered within the framework of the intra-nuclear cascade (INC)

model, as is usually done for antinucleon energies above several tens of $MeV$. Such an interaction also cannot be legitimately modeled using antineutron-nucleon cross sections. The approach used here to describe the interaction between the nucleus and the incoming slow antineutron resulting from the transformation is based on the integration of optical and cascade models. In the optical-cascade model, the *initial conditions* for the INC are formulated within the optical model. This approach was first applied in [28] to describe the annihilation of stopping antiprotons on nuclei when the antiproton is absorbed from the bound state made by the antiproton orbiting the atom. The same approach was used for the antineutron by L. A. Kondratyuk [24,25] in the discussion of future $n \to \bar{n}$ search experiments. The radial $(r)$ distribution of the absorption probability density $P_{abs}(r)$ is directly related to the radial nuclear density $\rho(r)$ and the radial wave function $\phi(r)$, and is derived from the wave equation for a slow antineutron:

$$P_{abs}(r) \sim 4\pi r^2 \rho(r)\,|\phi(r)|^2. \qquad (3)$$

This solution for a slow, plane wave antineutron incident on a $^{12}_{6}C$ nucleus was presented in great detail in [24,25]. In order to define the annihilation point in simulation, it is desirable to use a simple analytic function. Therefore, we approximate the solution $P_{abs}(r)$ obtained in [24,25] as a Gaussian function, with a maximum situated at $r = c + 1.2\,fm$, where $c$ is the radius of half density (with $c(^{12}_{6}C) = 2.0403\,fm$) with a width of $\sigma = 1\,fm$. This approximated function is presented in Fig. 1 as the solid orange curve with arbitrary units to demonstrate the penetration depth of the antineutron inside the nucleus.

The model assumes that the proton density within the nucleus $\rho(r)$ is described as an electrical charge distribution, as obtained in high-energy electron scattering experiments. The function $\rho(r)$ obeys a Woods-Saxon distribution:

$$\frac{\rho(r)}{\rho(0)} = \left[1 + e^{\frac{r-c}{a}}\right]^{-1}, \qquad (4)$$

where $a = 0.5227\,fm$ is the diffuseness parameter of the nucleus, and $c$ the radius of half density [29]. For practical reasons within the modeling process, the nucleus is split into seven concentric zones, within which the nucleon density is considered to be constant. Fig. 1 shows the density distribution of the nucleons for ${}^{12}_{6}C$, calculated by equation (4), along with a step approximation which divides the nucleus into seven zones of constant density. It is seen that although an antineutron penetrates more deeply compared to an antiproton (the dotted line in the Fig. 1), the absorption of the antineutron still occurs about the periphery of the nucleus. Since an antineutron would be strongly absorbed even within the diffuse periphery of the nuclear substance, another *eighth* zone with density $\rho_{out} = 0.001 \cdot \rho(7)$ is added which extends far beyond the nuclear envelope.

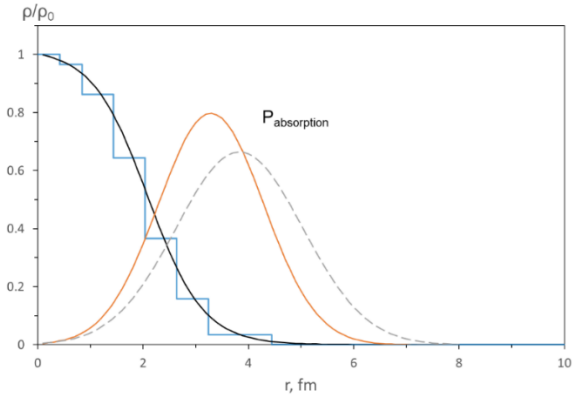

FIG. 1. Left: The radial distribution of the relative density of protons and neutrons throughout the ${}^{12}_{6}C$ nucleus (they are identical). The solid black line is a Woods-Saxon density distribution, while the blue step function is an approximation used to divide the nucleus into seven zones of constant density. Right: The radial dependence of the absorption probabilities $P_{abs}$ for the ${}^{12}_{6}C$ are shown for an antineutron (solid orange) and an antiproton (dashed grey) [28]. Note that the eighth zone extends from the end of zone seven at $r = 4.44\,fm$ to $r = 10\,fm$.

## B. Antineutron annihilation within the ${}^{40}_{18}Ar$ nucleus

For intranuclear $n \rightarrow \bar{n}$ transformation, the conversion of a bound neutron into an associated antineutron is significantly suppressed within the nuclear environment. The reason for this is the large difference between the values of the effective nuclear potential for the neutron and antineutron [5,10]. The question of the magnitude of the suppression of $n \rightarrow \bar{n}$ occurring within the nucleus has been the subject of in depth nuclear theoretical discussions for a number of years [10-13,30-32]. The transformation would be more probable for neutrons with lower binding energy, and the maximum of the antineutron wave-function is located beyond the nuclear radius [32]. In contrast to [18,23], for the correct description of the absorption process of the antineutron produced by $n \rightarrow \bar{n}$ within ${}^{40}_{18}Ar$, it is necessary to determine the radial dependence of the probability density of the transformation within the nucleus. This development will be included in our next publication, focused on ${}^{40}_{18}Ar$. However, for a first approximation of the annihilation process within ${}^{40}_{18}Ar$, the simulation outputs shown in *this* article are considered for the case when the transformation occurs with equal probability for all neutrons within *only* the peripheral zone of the nucleus. Thus, we plan to demonstrate the importance of the radial annihilation dependence. With regard to modeling the transformation in the ${}^{40}_{18}Ar$ nucleus, as discussed in the previous section, the difference between the absorption of the slow antineutron by the ${}^{12}_{6}C$ nucleus comes in the first stage only. As with the ${}^{12}_{6}C$ nucleus, the nucleon density distribution of the ${}^{40}_{18}Ar$ nucleus is described by expression (4) and is *approximated* as being divided into seven zones of constant density where it is assumed that the neutrons are distributed throughout the nucleus identically to protons.

## C. The nuclear model and nucleon momentum distribution

Within the INC model, the nucleus is considered to be a degenerate, free Fermi gas of nucleons, enclosed within a spherical potential well with a radius equal to the nuclear radius. Nucleons fill all energy levels of the potential well, from the lowest, when a nucleon can have the *largest* negative potential energy and ∼0 momentum, to the highest echelons of the Fermi level, where the nucleon moves with Fermi momentum $p_{FN}$, and is retained within the nucleus only because of the binding energy $\varepsilon$ (where $\varepsilon \approx 7\ MeV$ per nucleon).

In the interval $p\varepsilon[0, p_{FN}]$, the three-momentum of the nucleon can take all permissible values. The differential probability distribution of the nucleons with respect to the total momentum and kinetic energy [29] takes the form:

$$W(p) = \frac{3p^2}{p_{FN}^3}, \qquad p \leq p_{FN}, \qquad (5)$$

$$W(T) = \frac{3T^{\frac{1}{2}}}{2T_{FN}^{\frac{3}{2}}}, \qquad T \leq T_{FN}. \qquad (6)$$

Here, $T$ is the kinetic energy of a nucleon within the nucleus, and $T_{FN} = \frac{p_{FN}^2}{2m_N}$ represents the boundary Fermi kinetic energy, while $m_N$ is the mass of the nucleon. If the nucleons are distributed evenly throughout the spherical well having a radius $R = r_0 A^{\frac{1}{3}}$ (and where $r_0$ is 1.2-1.4 $fm$), then their Fermi momentum and energy are easily expressed in terms of the radius. Because every cell in phase space $d^3x\ d^3p$ contains a number of states

$$\frac{2s+1}{(2\pi\hbar)^3} d^3x\ d^3p \qquad (7)$$

($s$ is the spin of the nucleon) and the total number of protons or neutrons in the nucleus being equal to $n_N$, it then follows from the normalization condition that

$$\frac{2s+1}{(2\pi\hbar)^3} \int d^3x\ d^3p = \frac{Vp_{FN}^3}{3\pi^2\hbar^3} = n_N, \qquad (8)$$

and one finally gets that

$$p_{FN} = \hbar \left(\frac{3\pi^2 n_N}{V}\right)^{\frac{1}{3}}, \qquad (9)$$

$$T_{FN} = \frac{p_{FN}^2}{2m_N} = \frac{\hbar^2}{2m_N}\left(\frac{3\pi^2 n_N}{V}\right)^{\frac{2}{3}}, \qquad (10)$$

where $V = \frac{4}{3}\pi R^3$ is the volume of the nucleus, and $m_N$ remains the nucleon mass.

If the nucleus is subdivided into concentric spherical zones of constant density, the values of $p_{FN}$ and $T_{FN}$ for each zone are calculated similarly to equations (9) and (10), but with an $i$-th radius, and the density of the nucleons within this $i$-th zone. Fig. 2 shows the spatial distribution of the potential $V_N = -(T_{FN} + \varepsilon)$ for protons and neutrons in both $^{12}_{6}C$ and $^{40}_{18}Ar$ nuclei.

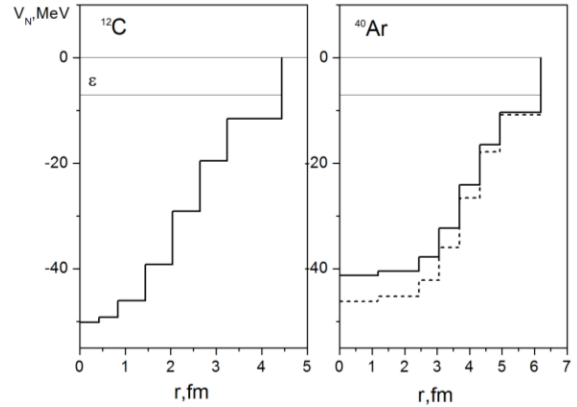

FIG. 2: The spatial distribution of the potential $V_N = -(T_{FN} + \varepsilon)$, with appropriate partitioning of the nucleus into seven zones for protons (solid histogram) and neutrons (dotted histogram) for both $^{12}_{6}C$ and $^{40}_{18}Ar$ nuclei (for $^{12}_{6}C$, the solid and dotted histograms lay atop one another). $\varepsilon$ is the average nuclear binding energy of 7 $MeV$ per nucleon.

The momentum distribution of the nucleons in individual zones will be the same as for a degenerate Fermi gas, and the probability of a nucleon to have momentum $p$ in the $i$th-zone will continue to be determined by (5), although corresponding to $i$-th-zone's boundary Fermi momentum value. Fig. 3 shows the momentum distributions of nucleons for both $^{12}_{6}C$ and $^{40}_{18}Ar$ nuclei, obtained by summing all the momentum distributions for all individual zones. From Figs. 2 and 3, we can see that the nucleons located in the central zone of the nucleus have the highest value of $T_{FN}$, and, accordingly, the maximum value of the Fermi momentum $p_{FN}$. Therefore, the contribution to the total momentum distribution from the nucleons located in the central ($i = 1$) zone gives the high-momentum part which extends up to $250 - 270 \ MeV/c$. Conversely, the nucleons located within the peripheral zone of the nucleus ($i = 7$) have momenta up to $80 - 100 \ MeV/c$. Moreover, the contribution to the overall momentum distribution of a particular zone is greater the more nucleons within it. Thus, in our model there is a correlation of the momentum with the density and, respectively, with the radius.

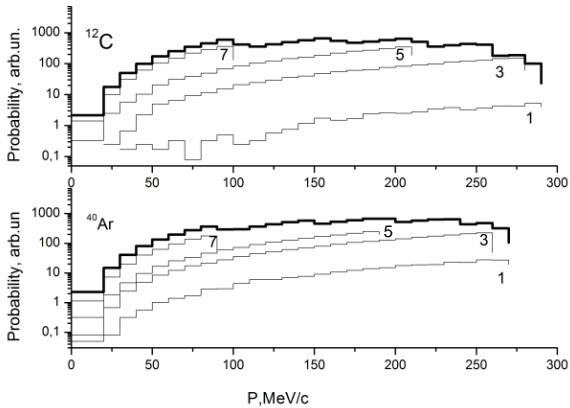

Fig. 3. The thick histogram shows the momentum distribution of intra-nuclear nucleons in both $^{12}_{6}C$ and $^{40}_{18}Ar$ nuclei, summed over all zones. The thin lines show histograms which correspond to contributions from individual zones of the nucleus to the total momentum distribution (only odd-numbered zone distributions are shown so that the picture is not indecipherable). The zone number $i$ is shown at the right of each histogram. Note the logarithmic scale of probability in arbitrary units.

Thus, for $^{12}_{6}C$ nuclei in the first stage, the nucleons are distributed within the nucleus according to the step density function (see Fig. 1). Next, according to the radial distribution of the antineutron absorption probability in the $^{12}_{6}C$ nucleus $P_{abs}(r)$ (also Fig. 1), the point of annihilation is taken randomly by Monte Carlo technique. The radius of this point determines the number of the zone in which the nucleon partner (neutron or proton) is located, and with which the antineutron annihilates.

The physics underlying the transformation within an $^{40}_{18}Ar$ nucleus is different, since the antineutron generated from $n \rightarrow \bar{n}$ is not extranuclear in nature, but instead depends entirely on the magnitude of the intra-nuclear binding energy as a function of radius. Generally, it is thought that the bound transformation should occur near the surface of the nucleus (possibly even *outside* the nuclear envelope [32]), we assume that an antineutron resulting from the transformation has kinetic energy $T_{\bar{n}} = T_n + \varepsilon$, and annihilates on the nearest nucleon neighbor within only the peripheral zone. Further, the simulation is done within the same scheme for both $^{12}_{6}C$ and $^{40}_{18}Ar$ nuclei. The annihilation partner has Fermi momentum randomly selected from the momentum distribution for a particular zone; then, the annihilation occurs.

### D. Annihilation model

Unlike papers [24,25], where a statistical model for nucleon-antinucleon annihilation into pions was used, the present paper uses a combined approach first proposed in [26]. The phenomena of $\bar{N}N$ annihilation can lead to the creation of many particles through many possible (at times $\sim 200$) exclusive reaction channels; many neutral particles may be present, which can make experimental study quite difficult. Experimental information for exclusive channels is known only for a small fraction of possible annihilation channels, and therefore a statistical model based on $SU(3)$ symmetry [33] has been chosen to describe the $\bar{N}N$ annihilation. Work to generalize the unitary-symmetric model for $\bar{N}N$ annihilations, along with the development of methods for calculating the characteristics of mesons produced from the annihilation, was performed by I. A. Pshenichnov [34]. According to the model, the $\bar{N}N$ annihilation allows for the production of between two and six intermediate particles. Given the estimates of the phase space volume at low momenta, the production of a larger number of intermediate particles is unlikely. Intermediate particles, such as $\pi, \rho, \omega$ and $\eta$ mesons, are all possible; the channels with strangeness production are not considered within this version of the model. This unitary-symmetric statistical model predicts 106 $\bar{p}p$ annihilation channels, and 88 $\bar{p}n$ annihilation channels, but this differs from experiments, which effectively measure only ~40 channels for $\bar{p}p$ and ~10 channels for $\bar{p}n$ annihilation. However, neither the statistical model, nor the experimental data, can provide a *complete* and *exclusive* description of the elementary nucleon-antinucleon annihilation processes. For this reason, *semi-empirical* Tables I and II of annihilation channels are employed for use in annihilation modeling. These are obtained as follows: First, all experimentally measured channels were included in Tables I and II. Then, by using isotopic relations, probabilities were found for those channels that have the same configurations but different particle charges. Finally, the predictions of the statistical model with $SU(3)$ symmetry were entered for the remaining intermediate channels. Sometimes the probabilities of intermediate channels measured in different experiments differ significantly. In this case, the data in the semi-empirical tables were corrected within experimental accuracies in order to describe the topological cross section for $\bar{p}p$ and $\bar{p}n$ in a consistent way. In our approach, a substantively large collection of experimental data was used: multi-particle topologies, inclusive spectra, topological pion cross sections, and branching ratios of various resonance channels. The following pages show the semi-empirical tables, with probabilities of various $\bar{p}p$ and $\bar{p}n$ annihilation channels included. In further modeling of $\bar{N}A$ interactions, it is considered that channels for $\bar{n}n$ are identical to $\bar{p}p$ channels, and that annihilation channels for $\bar{n}p$ are charge conjugated to $\bar{p}n$ channels.

Considering the laws of energy and momentum conservation for each annihilation, the procedure for simulating the characteristics of both the intermediate particles and their various decay products consists of the following: first, a single channel from the table is randomly selected via Monte Carlo technique as the initial state, with all necessary momenta of all annihilation particles determined according to the pertinent phase-space volume. This takes into account the Breit-Wigner mass distribution for meson resonances, while all pions have a mass value of $0.14 \frac{GeV}{c^2}$. The subsequent disintegration of unstable mesons is modeled according to experimentally known branching ratios. All major decay modes for meson resonances have been considered, such as in Table III.

All experimental data used for comparison with this annihilation model are described in great detail in [26]. However, there do exist more recent data obtained from LEAR by the Crystal Barrel [35] and OBELIX [36] Collaborations on some exclusive channels which show *somewhat* different branching ratios from those used by us. We plan to make a revision of the annihilation tables in the near future, taking into account *all* data. Below is a comparison of the simulation results and experimental data on $\bar{p}p$ annihilation at rest.

Table IV shows the average multiplicity of mesons formed in $\bar{p}p$ annihilations at rest. The simulation results are within the range of experimental uncertainties. From these simulation results, it follows that more than 35% of all pions have been formed by the decay of meson resonances. Fig. 4 shows the pion multiplicity distribution generated by $\bar{p}p$ annihilation, while Fig. 5 shows the charged pion momentum distribution. From considering Table IV and Figs. 4 and 5, it follows that the Monte Carlo and available experimental data are in general agreement with the main features of $\bar{p}p$ annihilation. In all, our annihilation model utilizes a complex series of tables with a much larger number of predicted and pertinent channels than [18,23]. As this approach demonstrates a good description of the experimental data for $\bar{p}p$ annihilation at rest, we believe that it is also adequate for an accurate description of $\bar{p}n$ annihilation at rest, and so can be implemented within our $\bar{n}A$ annihilation simulation.

TABLE I. Probability of intermediate states for $\bar{p}p$ annihilation at rest (%). Note that 1) indicates a probability attained from experiment; see references used in [26]. Note that 2) indicates that the probabilities are obtained from isotopic relations. The sum of all branching ratios is normalized to 100 percent.

| Channel | Probability (%) | Channel | Probability (%) | Channel | Probability (%) |
|---|---|---|---|---|---|
| $\eta\eta$ | 0.01  1) | $\eta\eta\pi^+\pi^-$ | 0.07 | $\pi^+\pi^+\pi^+\pi^-\pi^-\pi^-$ | 2.83 |
| $\eta\omega$ | 0.34  1) | $\eta\eta\pi^0\pi^0$ | 0.02 | $\pi^+\pi^+\pi^-\pi^-\pi^0\pi^0$ | 9.76 |
| $\omega\,\omega$ | 1.57  1) | $\eta\omega\pi^+\pi^-$ | 0.04 | $\pi^+\pi^-\pi^0\pi^0\pi^0\pi^0$ | 2.68 |
| $\pi^+\pi^-$ | 0.40  1) | $\eta\omega\pi^0\pi^0$ | 0.01 | $\pi^0\pi^0\pi^0\pi^0\pi^0\pi^0$ | 0.07 |
| $\pi^0\pi^0$ | 0.02  1) | $\pi^+\pi^-\pi^0\eta$ | 1.22 | $\pi^+\pi^+\pi^+\pi^-\pi^-\rho^-$ | 0.02 |
| $\pi^+\rho^-$ | 1.52  1) | $\pi^0\pi^0\pi^0\eta$ | 0.17 | $\pi^+\pi^+\pi^-\pi^-\pi^-\rho^+$ | 0.02 |
| $\pi^-\rho^+$ | 1.52  1) | $\pi^+\pi^-\pi^0\omega$ | 2.84 | $\pi^+\pi^+\pi^-\pi^-\pi^0\rho^0$ | 0.06 |
| $\pi^0\rho^0$ | 1.57  1) | $\pi^0\pi^0\pi^0\omega$ | 0.40 | $\pi^+\pi^+\pi^-\pi^0\pi^0\rho^-$ | 0.06 |
| $\rho^-\rho^+$ | 3.37  2) | $\pi^+\pi^-\rho^0\eta$ | 0.06 | $\pi^+\pi^-\pi^-\pi^0\pi^0\rho^+$ | 0.06 |
| $\rho^0\rho^0$ | 0.67  1) | $\pi^+\pi^0\rho^-\eta$ | 0.06 | $\pi^+\pi^-\pi^0\pi^0\pi^0\rho^0$ | 0.03 |
| $\pi^0\eta$ | 0.06  1) | $\pi^-\pi^0\rho^+\eta$ | 0.06 | $\pi^+\pi^0\pi^0\pi^0\pi^0\rho^-$ | 0.01 |
| $\pi^0\omega$ | 0.58  1) | $\pi^0\pi^0\rho^0\eta$ | 0.02 | $\pi^-\pi^0\pi^0\pi^0\pi^0\rho^+$ | 0.01 |
| $\rho^0\eta$ | 0.90  1) | $\pi^+\pi^-\pi^+\pi^-$ | 2.74 | $\pi^+\pi^+\pi^-\pi^-\pi^0\eta$ | 0.31 |
| $\rho^0\omega$ | 0.79  1) | $\pi^+\pi^-\pi^0\pi^0$ | 3.89 | $\pi^+\pi^-\pi^0\pi^0\pi^0\eta$ | 0.17 |
| $\pi^+\pi^-\pi^0$ | 2.34  1) | $\pi^0\pi^0\pi^0\pi^0$ | 0.21 | $\pi^0\pi^0\pi^0\pi^0\pi^0\eta$ | 0.01 |
| $\pi^0\pi^0\pi^0$ | 1.12  1) | $\pi^+\pi^+\pi^-\rho^-$ | 2.58  1) | $\pi^+\pi^+\pi^-\pi^-\pi^0\omega$ | 0.10 |
| $\pi^+\pi^-\rho^0$ | 2.02  1) | $\pi^+\pi^-\pi^-\rho^+$ | 2.58  1) | $\pi^+\pi^-\pi^0\pi^0\pi^0\omega$ | 0.06 |
| $\pi^+\pi^0\rho^-$ | 2.02  2) | $\pi^+\pi^-\pi^0\rho^0$ | 6.29  1) | $\eta\eta\eta$ | 0.0036 |
| $\pi^-\pi^0\rho^+$ | 2.02  2) | $\pi^+\pi^0\pi^0\rho^-$ | 5.05  2) | $\eta\eta\rho^0$ | 0.0002 |
| $\pi^0\pi^0\rho^0$ | 1.01  2) | $\pi^-\pi^0\pi^0\rho^+$ | 5.05  2) | $\omega\omega\pi^+\pi^-$ | 0.0002 |
| $\pi^+\rho^-\rho^0$ | 1.23 | $\pi^0\pi^0\pi^0\rho^0$ | 0.77  2) | $\omega\rho^0\pi^+\pi^-$ | 0.0005 |
| $\pi^-\rho^+\rho^0$ | 1.23 | $\pi^+\pi^+\pi^-\pi^-\pi^0$ | 2.61 | $\omega\rho^-\pi^+\pi^0$ | 0.0005 |
| $\pi^0\rho^+\rho^-$ | 1.23 | $\pi^+\pi^-\pi^0\pi^0\pi^0$ | 1.37 | $\omega\rho^+\pi^-\pi^0$ | 0.0005 |
| $\pi^0\rho^0\rho^0$ | 0.54 | $\pi^0\pi^0\pi^0\pi^0\pi^0$ | 0.07 | $\omega\rho^0\pi^0\pi^0$ | 0.0002 |
| $\pi^+\pi^-\eta$ | 1.50  1) | $\pi^+\pi^+\pi^-\pi^-\rho^0$ | 0.08 | $\rho^-\rho^-\pi^+\pi^+$ | 0.0003 |
| $\pi^+\pi^-\omega$ | 3.03  1) | $\pi^+\pi^+\pi^-\pi^0\rho^-$ | 0.16 | $\rho^0\rho^0\pi^0\pi^0$ | 0.0001 |
| $\pi^0\pi^0\omega$ | 0.79  2) | $\pi^+\pi^-\pi^-\pi^0\rho^+$ | 0.16 | $\rho^+\rho^-\pi^+\pi^-$ | 0.0011 |
| $\pi^+\rho^-\eta$ | 0.84 | $\pi^+\pi^-\pi^0\pi^0\rho^0$ | 0.12 | $\rho^0\rho^0\pi^+\pi^-$ | 0.0004 |
| $\pi^-\rho^+\eta$ | 0.84 | $\pi^+\pi^0\pi^0\pi^0\rho^-$ | 0.04 | $\rho^-\rho^0\pi^+\pi^0$ | 0.0008 |
| $\pi^0\rho^0\eta$ | 0.44 | $\pi^0\pi^0\pi^0\pi^0\rho^0$ | 0.01 | $\rho^+\rho^+\pi^-\pi^-$ | 0.0003 |
| $\pi^+\rho^-\omega$ | 1.10 | $\pi^+\pi^+\pi^-\pi^-\eta$ | 0.11  1) | $\rho^+\rho^0\pi^-\pi^0$ | 0.0008 |
| $\pi^-\rho^+\omega$ | 1.10 | $\pi^+\pi^-\pi^0\pi^0\eta$ | 0.22  2) | $\rho^+\rho^-\pi^0\pi^0$ | 0.0004 |
| $\pi^0\rho^0\omega$ | 0.57 | $\pi^0\pi^0\pi^0\pi^0\eta$ | 0.01  2) | $\pi^+\pi^-\pi^0\eta\eta$ | 0.0055 |
| $\eta\eta\pi^0$ | 0.11 | $\pi^+\pi^+\pi^-\pi^-\omega$ | 1.80  1) | $\pi^0\pi^0\pi^0\eta\eta$ | 0.0007 |
| $\eta\omega\pi^0$ | 0.30 | $\pi^+\pi^-\pi^0\pi^0\omega$ | 2.58  2) | | |
| $\omega\omega\pi^0$ | 0.37 | $\pi^0\pi^0\pi^0\pi^0\omega$ | 0.10  2) | | |

TABLE II. Probability of intermediate states for $\bar{p}n$ annihilation at rest (%). Similarly note that 1) indicates a probability attained from experiment; see references used in [26]. Note that 2) indicates that the probabilities are obtained from isotopic relations. The sum of all branching ratios is normalized to 100 percent.

| Channel | Probability (%) | | Channel | Probability (%) | | Channel | Probability (%) |
|---|---|---|---|---|---|---|---|
| $\pi^-\pi^0$ | 0.49 | 1) | $\pi^+\pi^-\pi^-\omega$ | 10.52 | 1) | $\pi^+\pi^+\pi^-\pi^-\pi^0\rho^-$ | 0.07 |
| $\pi^-\omega$ | 0.48 | 1) | $\pi^-\pi^0\pi^0\omega$ | 7.01 | 2) | $\pi^+\pi^-\pi^-\pi^-\pi^0\rho^+$ | 0.05 |
| $\pi^-\rho^0$ | 0.47 | 1) | $\pi^+\pi^-\rho^-\eta$ | 0.08 | | $\pi^+\pi^-\pi^-\pi^0\pi^0\rho^0$ | 0.06 |
| $\pi^0\rho^-$ | 0.47 | 2) | $\pi^-\pi^-\rho^+\eta$ | 0.05 | | $\pi^+\pi^-\pi^0\pi^0\pi^0\rho^-$ | 0.03 |
| $\rho^-\rho^0$ | 3.51 | 2) | $\pi^-\pi^0\rho^0\eta$ | 0.06 | | $\pi^-\pi^-\pi^0\pi^0\pi^0\rho^+$ | 0.02 |
| $\pi^-\eta$ | 0.29 | 1) | $\pi^0\pi^0\rho^-\eta$ | 0.02 | | $\pi^-\pi^0\pi^0\pi^0\pi^0\rho^0$ | 0.01 |
| $\rho^-\eta$ | 2.27 | | $\pi^+\pi^-\pi^-\pi^0$ | 5.51 | | $\pi^+\pi^-\pi^-\pi^-\eta$ | 0.14 |
| $\rho^-\omega$ | 3.51 | 2) | $\pi^-\pi^0\pi^0\pi^0$ | 1.38 | | $\pi^+\pi^-\pi^-\pi^0\pi^0\eta$ | 0.30 |
| $\pi^+\pi^-\pi^-$ | 2.86 | | $\pi^+\pi^-\pi^-\rho^0$ | 0.99 | | $\pi^-\pi^0\pi^0\pi^0\pi^0\eta$ | 0.05 |
| $\pi^-\pi^0\pi^0$ | 1.90 | | $\pi^+\pi^-\pi^0\rho^-$ | 1.97 | | $\pi^+\pi^+\pi^-\pi^-\pi^-\omega$ | 0.05 |
| $\pi^+\pi^-\rho^-$ | 3.62 | 1) | $\pi^-\pi^-\pi^0\rho^+$ | 0.99 | | $\pi^+\pi^-\pi^-\pi^0\pi^0\omega$ | 0.09 |
| $\pi^-\pi^-\rho^+$ | 0.58 | 1) | $\pi^-\pi^0\pi^0\rho^0$ | 0.75 | | $\pi^-\pi^0\pi^0\pi^0\pi^0\omega$ | 0.01 |
| $\pi^-\pi^0\rho^0$ | 5.61 | 2) | $\pi^0\pi^0\pi^0\rho^-$ | 0.25 | | $\eta\eta\rho^-$ | 0.0003 |
| $\pi^0\pi^0\rho^-$ | 3.51 | 2) | $\pi^+\pi^+\pi^-\pi^-\pi^-$ | 1.24 | | $\omega\omega\pi^-\pi^0$ | 0.0002 |
| $\pi^+\rho^-\rho^-$ | 1.04 | | $\pi^+\pi^-\pi^-\pi^0\pi^0$ | 2.72 | | $\omega\rho^-\pi^+\pi^-$ | 0.0008 |
| $\pi^-\rho^+\rho^-$ | 2.09 | | $\pi^-\pi^0\pi^0\pi^0\pi^0$ | 0.37 | | $\omega\rho^+\pi^-\pi^-$ | 0.0004 |
| $\pi^-\rho^0\rho^0$ | 0.70 | | $\pi^+\pi^+\pi^-\pi^-\rho^-$ | 0.12 | | $\omega\rho^0\pi^-\pi^0$ | 0.0005 |
| $\pi^0\rho^-\rho^0$ | 1.39 | | $\pi^+\pi^-\pi^-\pi^-\rho^+$ | 0.08 | | $\omega\rho^-\pi^0\pi^0$ | 0.0003 |
| $\pi^-\pi^0\eta$ | 1.23 | | $\pi^+\pi^-\pi^-\pi^0\rho^0$ | 0.16 | | $\rho^-\rho^0\pi^+\pi^-$ | 0.0011 |
| $\pi^-\pi^0\omega$ | 5.05 | | $\pi^+\pi^-\pi^0\pi^0\rho^-$ | 0.16 | | $\rho^-\rho^-\pi^+\pi^0$ | 0.0005 |
| $\pi^0\rho^-\eta$ | 0.78 | | $\pi^-\pi^-\pi^0\pi^0\rho^+$ | 0.08 | | $\rho^+\rho^0\pi^-\pi^-$ | 0.0005 |
| $\pi^-\rho^0\eta$ | 0.78 | | $\pi^-\pi^0\pi^0\pi^0\rho^0$ | 0.05 | | $\rho^-\rho^+\pi^0\pi^-$ | 0.0011 |
| $\pi^-\rho^0\omega$ | 1.03 | | $\pi^0\pi^0\pi^0\pi^0\rho^-$ | 0.01 | | $\rho^0\rho^0\pi^0\pi^-$ | 0.0004 |
| $\pi^0\rho^-\omega$ | 1.03 | | $\pi^+\pi^-\pi^-\pi^0\eta$ | 0.37 | | $\rho^-\rho^0\pi^0\pi^0$ | 0.0004 |
| $\eta\eta\pi^-$ | 0.21 | | $\pi^-\pi^0\pi^0\pi^0\eta$ | 0.09 | | $\pi^+\pi^-\pi^-\eta\eta$ | 0.0042 |
| $\pi^-\omega\eta$ | 0.60 | | $\pi^+\pi^-\pi^-\pi^0\omega$ | 0.40 | | $\pi^-\pi^0\pi^0\eta\eta$ | 0.0028 |
| $\pi^-\omega\omega$ | 0.71 | | $\pi^-\pi^0\pi^0\pi^0\omega$ | 0.09 | | | |
| $\eta\eta\pi^-\pi^0$ | 0.06 | | $\pi^+\pi^+\pi^-\pi^-\pi^-\pi^0$ | 8.33 | | | |
| $\eta\omega\pi^-\pi^0$ | 0.03 | | $\pi^+\pi^+\pi^-\pi^-\pi^0\pi^0$ | 6.67 | | | |
| $\pi^+\pi^-\pi^-\eta$ | 1.00 | | $\pi^-\pi^0\pi^0\pi^0\pi^0\pi^0$ | 0.56 | | | |
| $\pi^-\pi^0\pi^0\eta$ | 0.67 | | $\pi^+\pi^+\pi^-\pi^-\rho^0$ | 0.02 | | | |

TABLE III. Pertinent decay branching ratios of intermediate resonance particles shown in %.

| Channel | Probability (%) | Channel | Probability (%) | Channel | Probability (%) |
|---|---|---|---|---|---|
| $\eta \to 2\gamma$ | 39.3 | $\omega \to \pi^+\pi^-\pi^0$ | 89.0 | $\rho^+ \to \pi^+\pi^0$ | 100 |
| $\eta \to 3\pi^0$ | 32.1 | $\omega \to \pi^0\gamma$ | 8.7 | $\rho^- \to \pi^-\pi^0$ | 100 |
| $\eta \to \pi^+\pi^-\pi^0$ | 23.7 | $\omega \to \pi^+\pi^-$ | 2.3 | $\rho^0 \to \pi^+\pi^-$ | 100 |
| $\eta \to \pi^+\pi^-\gamma$ | 4.9 | | | | |

TABLE IV. Meson multiplicities for simulated and experimental $\bar{p}p$ annihilations, shown in absolute particle counts.

| Multiplicity | Simulated $\bar{p}p$ | Experimental $\bar{p}p$ |
|---|---|---|
| $M(\pi)$ | 4.910 | 4.98±0.35   [37] <br> 4.94±0.14   [38] |
| $M(\pi^{\pm})$ | 3.110 | 3.14±0.28   [37] <br> 3.05±0.04   [37] <br> 3.04±0.08   [38] |
| $M(\pi^0)$ | 1.800 | 1.83±0.21   [37] <br> 1.93±0.12   [37] <br> 1.90±0.12   [38] |
| $M(\eta)$ | 0.091 | 0.10±0.09   [39] <br> 0.0698±0.0079   [37] |
| $M(\omega)$ | 0.205 | 0.28±0.16,   [39] <br> 0.22±0.01   [40] |
| $M(\rho^+)$ | 0.189 | --- |
| $M(\rho^-)$ | 0.191 | --- |
| $M(\rho^0)$ | 0.193 | 0.26±0.01   [40] |
| $M(\pi)_{decay}^{from}$ | 1.908 | --- |
| $M(\pi^+)_{decay}^{from}$ | 0.606 | --- |
| $M(\pi^-)_{decay}^{from}$ | 0.606 | --- |
| $M(\pi^0)_{decay}^{from}$ | 0.695 | --- |

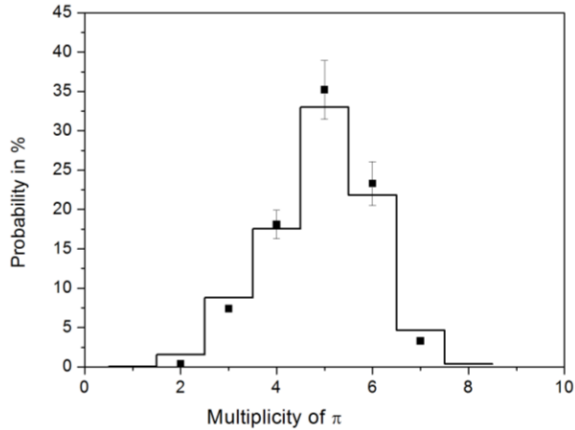

FIG. 4. The pion multiplicity distribution for $\bar{p}p$ annihilation at rest (taking into account the decay of meson resonances). The solid histogram shows the model, with the points showing experimental data [37].

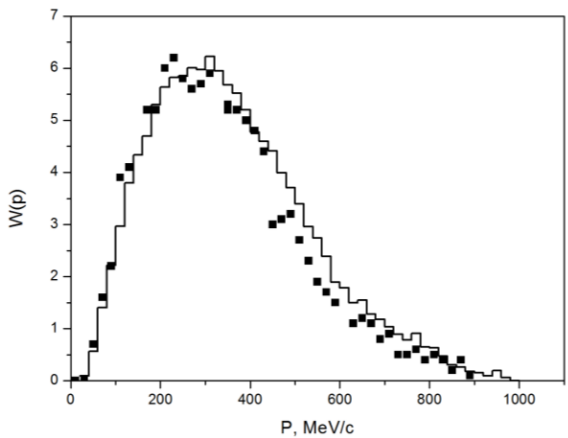

FIG. 5. The momentum distribution of charged pions produced in $\bar{p}p$ annihilation at rest (taking into account the decay of meson resonances). The solid histogram shows the model, with the points showing experimental data [41].

### E. The Intra-nuclear Cascade (INC) Model

Inelastic nuclear interactions are clearly statistical in nature, as they can be realized in many possible states. A statistical approach is key to describing such systems, and replaces the evolution of a system's wave function with the description of the evolution of an ensemble of the many possible states of the system. There are two dramatically different stages of a deeply inelastic interaction: 1) a fast, out-of-equilibrium stage in which energy is redistributed between the various degrees of freedom within the nucleus as a finite open system, and 2) the slow equilibrium stage of the decay of the thermalized residual nuclei.

The INC model is a phenomenological model describing the out-of-equilibrium stage of inelastic interactions and operates with the notion of the probability of a nuclear system being in a given state. Transitions between different states are caused by two-body interactions, which lead to secondary particles exiting the nucleus, dissipating the excitation energy in the process. However, this phenomenological model is linked to fundamental microscopic theory. It was shown in [42] that it is possible to transform a non-stationary Schrödinger equation for a many body system into kinetic equations, if large energy (and so short time) wave packet formulations are used. To explain, if the duration of the wave packet's individual collisions are shorter than the interval of time between consecutive collisions, then the amplitudes of these collisions will not interfere. This condition is essentially analogous to the condition of a free gas approximation: $\tau_0 < \tau_{FP}$, where $\tau_0$ is the duration time of the collision, and $\tau_{FP}$ is the mean-free-path time. This condition allows for the consideration of a particle's motion as in a dilute gas with independent particle motion on free path trajectories perturbed by binary collisions. Under these conditions, in a quasi-classical way, one can use the local momentum approximation by assigning a particle a momentum $\vec{P}(\vec{r})$ between consecutive collisions. In this case, the quantum kinetic equation is transformed into a kinetic equation of Boltzmann-type describing the transport of particles within nuclear media; this differs from the conventional Boltzmann equation only by accounting for the Pauli exclusion principle. Thus, the INC model is a numerical solution of the quasi-classical kinetic equation of motion for a multi-particle distribution function using the Monte Carlo method.

We will now focus our discussion on the scope of the INC model and the possibility of generalizing its use, such as in the event of the absorption of a slow antineutron. The principles underlying the model are altogether justified if the following conditions are met [29,42,43]:

a) The wavelengths, $\lambda$, of the majority of moving particles should be less than the mean distance between nucleons within the nucleus, i.e. $\lambda < \Delta$, where

$$\Delta \approx \left[ \frac{4\pi R^3}{3A} \right]^{\frac{1}{3}} \approx r_0 \approx 1.3 \; fm.$$

In this case, the system acquires quasi-classical characteristics, and one can speak of the trajectories of particles and two-body interactions within the nucleus. For individual nucleons, this corresponds to an energy of more than tens of $MeV$. Of course, this condition cannot be met in the case of a slow antineutron, and therefore, its absorption is described in the framework of the optical model.

b) The interaction time should be less than the time between successive interactions $\tau_{int} \leq \tau_{FP}$, where $\tau_{int} \approx \frac{r_N}{c} \approx 10^{-23}s$, and $r_N$ is the nucleon radius. The mean-free-path-length time is

$$\tau_{FP} = \frac{l}{c} = \frac{1}{\rho \sigma c} \approx \frac{4\pi R^3}{3A\sigma c} \approx \frac{3 \times 10^{-22}}{\sigma} s$$

($\sigma$ is the cross section in $mB$). This requirement is equivalent to the condition of requiring sufficiently small cross sections of elementary interactions and proves problematic for pions produced from the annihilation and lying within the energy range of the $\Delta$-resonance, where $\sigma > 100 \; mB$. However, it should be kept in mind that the effective mean-free-path-length within the nucleus is *increased* by the Pauli exclusion principle; secondarily, because the uptake of the antineutron is predominately on the periphery of the nucleus, where the nuclear density is low and the distance between the nucleons large, one can expect that the INC model would work in this case. Never-the-less, the comparison of the simulation results with experimental data is the main criterion for the applicability of the model.

The standard INC model is based on a numerical solution of the kinetic equation using a linearized approximation, which implies that the density of the media *does not change* in the development of the cascade, i.e. $N_c \ll A_t$ (where $N_c$ is the number of cascading particles, and $A_t$ is the number of nucleons making up the target nucleus). Such an approximation is violated in the case of multi-pion production in $pA$ and $\pi A$ interactions at $E_{p,\pi} \geq 3 - 5 \; GeV$, and also in the case of annihilation, especially when considering light nuclei such as $^{12}_{6}C$. A version of the model, which takes into account the effect of a local reduction in nuclear density, was first proposed in [43]. This version of the model considers the nucleus as consisting of separate nucleons, the position of their centers computed by Monte Carlo method according to the prescribed density distribution $\rho(R)$ such that the distance between their centers is no less than $2r_c$, where $r_c = 0.2 \; fm$ is the nucleon core radius. A cascading particle may interact with any intra-nuclear nucleon which lies inside the cylinder of diameter $2r_{int} + \lambda$ extending along the particle's velocity vector (here, $r_{int}$ is the interaction radius, while $\lambda$ is the deBroglie wavelength of the particle). The $r_{int}$ is a parameter of the model and is chosen for better agreement with the experimental data. The key point to understand here is the ability to determine the probability of the cascading particle interacting with another constituent nucleon. We now consider this process in more detail.

Within the standard cascade model, the randomly chosen interaction point is computed from a Poisson distribution for the mean-free-path-length. In this case, the probability $\omega(k)$ of the particle experiencing $k$ collisions along the path-length $L$ in media with density $\rho$, where the particle has a total cross section $\sigma$, is defined as:

$$\omega(k) = e^{-\rho \sigma L} \frac{(\rho \sigma L)^k}{k!}. \qquad (11)$$

If on the path-length $L$ there lie $n$ individual particle centers, each has an equal collision probability $p$ for the particle to collide on $k$ of $n$ centers and $q = 1 - p$. This probability is described by a binomial distribution:

$$\omega(k, n, p) = \frac{n!}{k!\,(n-k)!}\, p^k q^{n-k}. \qquad (12)$$

From the Poisson distribution (11), it follows directly that the probability of a particle experiencing *no collisions* along $L$ is simply:

$$\omega(0) = e^{-\rho\sigma L}. \qquad (13)$$

The same probability for this process can be obtained from the binomial distribution in (20):

$$\omega(0, n, p) = (1-p)^n = q^n. \qquad (14)$$

If one takes $\omega(0) = \omega(0, n, p)$, and when considering that $n = \rho\pi L(r_{int} + \lambda/2)^2$, then:

$$q = 1 - p = \exp\left[-\frac{\rho\sigma L}{n}\right]$$
$$= \exp[-\sigma/\pi(r_{int} + \lambda/2)^2]. \qquad (15)$$

An essential feature of present version of the INC model is the fact that after interactions occur inside the nucleus, the nucleon is considered to be cascade particle and not a constituent part of the nuclear system. Thus, a reduction in nuclear density takes place during the cascade development. In order to describe the evolution of the cascade and the decays of unstable meson resonances over time, an explicit time-coordinate has been incorporated into the model.

So, one can summarize the physical considerations that underlie the INC model as follows:

- The nuclear target is a degenerate Fermi gas of protons and neutrons within a spherical potential well with a diffuse nuclear boundary. The real nuclear potentials for nucleons $(V_N)$, antinucleons $(V_{\bar{N}})$, and mesons $(V_\pi, V_\eta, V_\omega)$ effectively takes into account the influence on the particle of all intra-nuclear nucleons. The depth of the potential well for the antinucleon and mesons within the nucleus remains a free-parameter of the model. Recognizing that the annihilation process usually occurs on the periphery of the nucleus, a good approximation for this is considered to be $V_{\bar{N}} \approx 0$ and $V_{\pi,\eta,\omega} \approx 0$. In the future, a detailed study is planned to focus on the influence of these potentials on the simulation output.
- Hadrons involved in collisions are treated as classical particles. A hadron can initiate a cascade of consecutive, independent collisions upon nucleons within the target nucleus. The interactions between cascading particles are not taken into account.
- The cross sections of hadron-nucleon interactions are considered within the nucleus to be identical to those in vacuum, except that Pauli's exclusion principle explicitly prohibits transitions of cascade nucleons into states already occupied by other nucleons.

| | | |
|---|---|---|
| $NN \to NN$ | $NN \to \pi NN$ | $NN \to i\pi\, NN \quad (i \geq 2)$ |
| $\pi N \to \pi N$ | $\pi + (NN) \to NN$ | $\pi N \to \pi\pi N$ |

$$(16)$$

Elementary processes, such as those seen in the channels (16) shown above, are described by empirical approximations from analysis of experimental data on $NN$ and $\pi N$ interactions at kinetic energies $T < 20\,GeV$ [29,43]

Now consider some of the features of the INC model related to the introduction of unstable meson resonances into the model. Modeling annihilation with meson resonances (i.e. $\bar{N}N \rightarrow i\pi + j\rho + n\eta + n\omega$) was described in the preceding section. It is assumed within the model that $\rho$-mesons produced by annihilation decay quickly enough to avoid interacting with any intra-nuclear nucleons. In contrast, $\omega$-mesons produced by annihilation can both interact with other intra-nuclear nucleons *and* decay within or outside the nucleus. The competition between the decay of the $\omega$-meson and its interaction with intra-nuclear nucleons is determined by the expression for the mean-free-path:

$$\frac{1}{\lambda} = \frac{1}{\lambda_{dec}} + \frac{1}{\lambda_{int}}, \qquad (17)$$

where $\lambda_{int} = (\rho_n \sigma_{\omega N}^{tot})^{-1}$, $\lambda_{dec} = \gamma\beta(h\Gamma_\omega)^{-1}$, $\rho_n$ is the nuclear density, and $\gamma$ is the Lorentz factor. The mean lifetime of the $\eta$-meson is large enough for the particle to be considered stable within the nucleus, which can then decay upon exit. The model uses the experimentally measured decay modes of the meson resonances described above. When the annihilation products are allowed to disintegrate, their three-body decay is simulated by evaluation of the permissible phase-space volume.

To accommodate the passage of $\eta$-and-$\omega$-mesons through nuclear material, in addition to channels listed in (16), the following interactions are also considered:

| | | | | |
|---|---|---|---|---|
| $\eta N \rightarrow \eta N$ | $\eta N \rightarrow \pi N$ | $\eta N \rightarrow \pi\pi N$ | $\eta + (NN) \rightarrow NN$ | $\eta + (NN) \rightarrow \pi NN$ |
| $\omega N \rightarrow \omega N$ | $\omega N \rightarrow \pi N$ | $\omega N \rightarrow \pi\pi N$ | $\omega + (NN) \rightarrow NN$ | $\omega + (NN) \rightarrow \pi NN$ |

$$(18)$$

Along with the creation of $\eta$- and $\omega$-mesons by annihilation, the model also accounts for the creation of mesons through interactions between annihilation pions and nucleons, such as

$$\pi N \rightarrow \eta N \qquad \pi N \rightarrow \omega N \qquad (19)$$

For cross sections of reactions in (18), estimates given in [26] were employed. For those few reactions shown in (19), experimental cross sections were taken from compilation [44]. As these interactions are considered at relatively low energy, the angular distributions for reactions shown in (18) and (19) are assumed to be isotropic in the center of mass of the system. Reactions with three particles in the final state are simulated via their pertinent phase-space volume.

### F. De-excitation of the residual nucleus

For inelastic nuclear reactions, after the rapid stage of the intra-nuclear cascade ($\tau_{cas} \simeq \tau_0$) and once statistical equilibrium ($\tau_{eq} \cong (5-10)\tau_0$) is established inside the residual nucleus, a slow stage begins ($\tau_{ev} \gg \tau_0$) involving the disintegration of the highly excited residual nucleus (note that $\tau_0 \leq 10^{-22}s$, which is the average time required for a particle to pass completely through the nucleus). The INC model is able to describe the dissipation of energy throughout the nucleus. At the end of the cascade stage, the nuclear degenerate Fermi gas contains a number of "holes" $N_h$, which is equal to the number of collisions of cascade particles with nucleons within the nucleus. Also, there exists some number of excited particles $N_p$, which is equal to the number of slow cascade nucleons trapped by the nuclear potential well. The excitation energy of the residual nucleus $E^*$, is the sum of the energy of all such quasiparticles calculated from the Fermi energies $\varepsilon_i$:

$$E^* = \sum_{i=1}^{N_h} \varepsilon_i^h + \sum_{j=1}^{N_p} \varepsilon_j^p. \qquad (20)$$

The resulting residual nuclei have a broad distribution on the excitation energies $E^*$, momenta, masses, and charges. The INC model correctly accounts for the fluctuations of the cascade particles, and reliably defines the entire set of characteristics for residual nuclei.

The de-excitation mechanism for a residual nucleus is determined from the accumulated excitation energy of the nucleus [45]. Under low excitation energies (where $E^* \leq 2 - 3 \frac{MeV}{nucleon}$), the primary de-excitation mechanism is the consecutive emission (evaporation) of particles from the compound nucleus [46]. When the excitation energy of the nucleus is approaching the total binding energy (where $E^* \geq 5 \frac{MeV}{nucl.}$), the prevalent mechanism is explosive decay [47]. For intermediate energies, both mechanisms coexist.

## III.    COMPARISON WITH EXPERIMENT

The optical-cascade model described throughout this work has been used to analyze experimental data taken from antiproton annihilation at rest on $^{12}_{6}C$ target nuclei. Table V shows the average multiplicity of emitted pions and protons. Experimental data on average pion multiplicities (values of which are shown at the bottom of the $\bar{p}C$ row) are taken from [38]. The final column of the table indicates the average energy of pions and photons (resulting from the decay of $\eta$- and $\omega$-mesons) emitted from the nucleus. Calculated values for the average multiplicities of pions (values of which are shown at the top of the $\bar{p}C$ row) are within accuracies of the experimental data. Since the antiproton primarily annihilates on the surface of the nucleus, most of the mesons produced fly out of the nucleus *without* any interaction. In the case of a light nucleus such as $^{12}_{6}C$, the effect of absorption of annihilation mesons is not large and the average multiplicity of pions emitted appear to be quite similar to the multiplicity of pions in $\bar{p}p$ annihilation (4.910). For comparison, Table V also shows results which simulate the annihilation of a slow antineutron on a $^{12}_{6}C$ nucleus. The comparison shows that the average pion multiplicity for an $\bar{n}C$ annihilation is somewhat *lower*, and that the average multiplicity for exiting nucleons slightly *higher* than the case of a stopped antiproton. This is due to the fact that the antineutron penetrates more deeply into the nucleus (seen in the solid line shown in Fig. 1) compared to an antiproton (seen in the dashed line shown in Fig. 1), and so there are more intra-nuclear interactions between annihilation mesons and constituent nucleons. Thus, the number of mesons emitted from the nucleus and their total energy $E_{tot}$ are *reduced*, while instead the number of nucleons that were kicked from the original nucleus during the fast cascading stage (and then emitted from the nucleus during the de-excitation process) is *increased*. In the case of peripheral annihilation of an antineutron on $^{40}_{18}Ar$, the pions are almost entirely free to leave the nucleus, increasing the value of $E_{tot}$, and so the number of emitted nucleons is significantly lower than $\bar{n}C$ annihilation. Note here that the calculation completed for $^{40}_{18}Ar$ is made in a *very rough* approximation with respect to the annihilation radius; this property requires further detailed investigation.

TABLE V. The average outgoing particle multiplicities emitted after $\bar{p}C$, $\bar{n}C$ and $\bar{n}Ar$ annihilation and all decays. Experimental data taken from [38] is used as a comparison for $\bar{p}C$. In the first row (calculation for $\bar{p}C$), an option of the model with the antineutron potential was used.

| | Type | $M_\pi$ | $M_{\pi^+}$ | $M_{\pi^-}$ | $M_{\pi^0}$ | $M_p$ | $M_n$ | $E_{tot}$ (MeV) |
|---|---|---|---|---|---|---|---|---|
| $\bar{p}C$ | *calculation* | 4.557 | 1.208 | 1.634 | 1.715 | 1.138 | 1.209 | 1736 |
| | *experiment* | $4.57 \pm 0.15$ | $1.25 \pm 0.06$ | $1.59 \pm 0.09$ | $1.73 + 0.10$ | --- | --- | $1758 \pm 59$ |
| $\bar{n}C$ | *calculation* | 4.451 | 1.558 | 1.182 | 1.712 | 1.543 | 1.317 | 1679 |
| $\bar{n}Ar$ | *calculation* | 4.599 | 1.651 | 1.267 | 1.680 | 0.727 | 0.804 | 1751 |

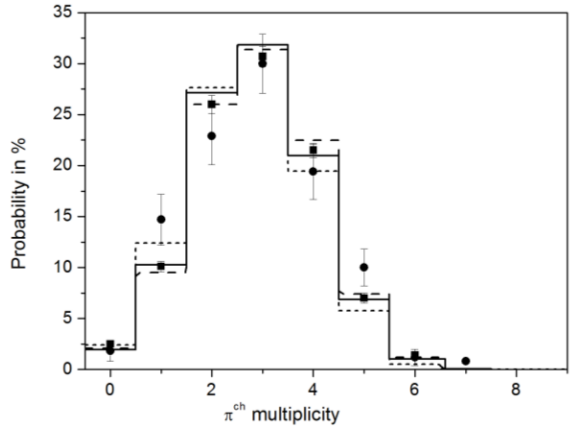

FIG. 6. The probability (%) of formation of a given number of charged pions in antinucleon-nuclei annihilation. The solid histogram shows $\bar{p}C$. Experimental data: circles-[48], squares-[49]. The dotted histogram shows an $\bar{n}C$ simulation; the dashed histogram shows an $\bar{n}Ar$ simulation.

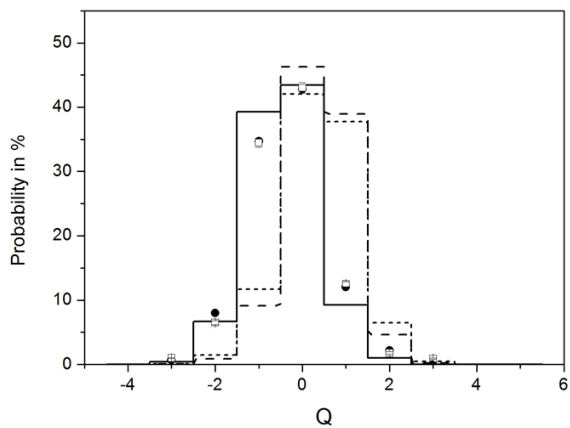

FIG. 7. The probability (%) of particular values of total charge $Q$ carried out by pions emitted from the nucleus. The solid histogram shows a $\bar{p}C$ calculation. Experimental data: open squares-[50], circles-[51]. The dotted histogram shows an $\bar{n}C$ simulation; the dashed histogram shows an $\bar{n}Ar$ simulation.

Now, consider and compare the Monte Carlo calculation to other available experimental data and features for $\bar{p}C$ annihilation at rest. Fig. 6 shows the charged pion multiplicity distribution emitted from the nucleus due to $\bar{p}C$ annihilation (shown as the solid histogram with points), $\bar{n}C$ (shown as the dotted histogram), and $\bar{n}Ar$ (the dashed histogram). As was expected, the differences in these distributions, as with the mean number of emitted pions, are not significant, although there appears to be some bias towards a smaller number of pions for $\bar{n}C$ and a larger number for $\bar{n}Ar$.

Fig. 7 shows the distribution by number of events with the charge $Q$ carried out by pions. For the $\bar{p}C$ annihilation the maxima of the distribution are $Q = -1$ and $Q = 0$, which *practically* corresponds to mesons exiting the nucleus without any interaction with nucleons. The optical-cascade model demonstrates good agreement with the experimental data. In the case of an annihilation with an $\bar{n}$, the distribution has a maximum $Q$ that is shifted to $Q = 0$ and $Q = +1$, respectively. In the case of a peripheral annihilation for $\bar{n}Ar$, the distribution has a narrower maximum than $\bar{n}C$.

Fig. 8 shows the distribution of the number of emitted protons. The analysis of experimental data and simulation results show that a significant number of events (from ~40% for $\bar{n}C$, to ~60% for $\bar{n}Ar$) do not have *any* exiting protons.

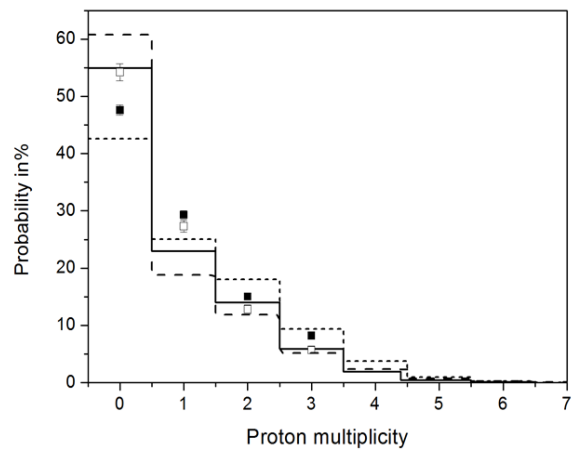

FIG. 8. The probability (%) of the events with a given number of exiting protons. The solid histogram shows a $\bar{p}C$ calculation. Experimental data: solid squares-[49], open squares-[50]. The dotted histogram shows an $\bar{n}C$ simulation. The dashed histogram shows an $\bar{n}Ar$ simulation.

Fig. 9 shows the momentum distribution for $\pi^+$ exiting the nucleus, which is rather similar to the momentum distribution of pions created by $\bar{p}p$ annihilation (as seen in Fig. 5). To understand the uncertainty of the model, calculations were done 1) without any nuclear potential for the antineutron, and, as an option, 2) with a model where the antineutron nuclear potential is introduced similarly to [26]. For mesons propagating inside the nucleus, we have not assumed any nuclear potentials. Both model calculations are presented in Fig. 9, and show rather good agreement with experimental data, although there is some exaggerated absorption behavior corresponding to the $\Delta$-resonance region ($\sim 260 \frac{MeV}{c}$). The difference between experimental measurements appears to be of the same order as the uncertainty in the calculation. Never-the-less, in the near future, we plan to study this question in more detail.

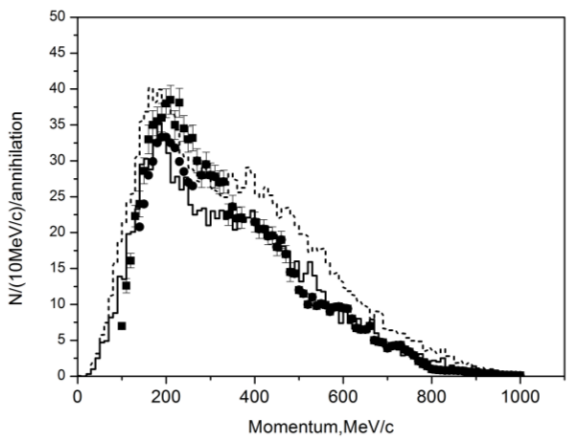

FIG. 9. The $\pi^+$ momentum distribution is shown for antiproton annihilation at rest on $^{12}_{6}C$ nuclei. The points show experimental data from [38,52]. The histograms show calculations, where the solid line shows an option of the model with a nuclear potential for the antineutron, and the dashed line is the calculation done without this potential. The nuclear potential for annihilation mesons inside the nucleus is assumed to be zero.

Fig. 10 shows the energy spectrum of protons exiting the nucleus from $\bar{p}C$ annihilation at rest. In the low energy regime (up to $50\,MeV$), evaporative protons provide a significant contribution to the spectrum. The model again shows good agreement with the available experimental data.

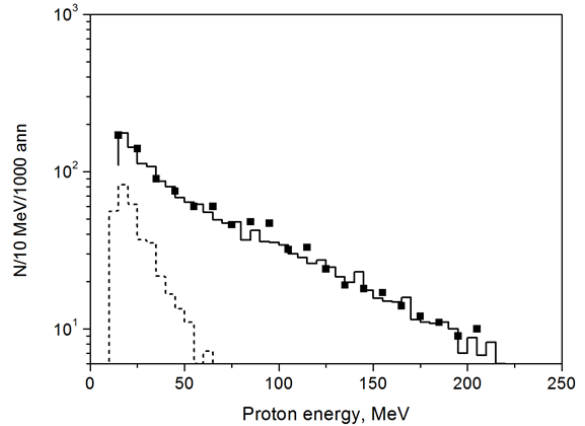

FIG. 10. The exiting proton kinetic energy spectrum due to antiproton annihilation at rest on $^{12}_{6}C$ nuclei. The solid histogram shows the simulation result. The dotted histogram shows the contribution which evaporative protons impart to the whole distribution. The points show the experimental data taken in [53].

From the comparisons made above between the simulation results of the optical-cascade model and experimental data of $\bar{p}C$ annihilation at rest, it follows that the model as a whole describes experiments well, thus accurately reflecting the dynamics of the annihilation process and the propagation of annihilation mesons throughout the nucleus.

## IV. $\bar{n}C$ ANNIHILATION GENERATOR VALIDATION

Colleagues within the ESS NNBar Collaboration have tested the model through its corresponding event generator comprehensively. The event generator outputs an annihilation point within $^{12}_{6}C$, annihilation product particle identities, energies, momenta, etc. These particles and variables are tracked as outputs and saved to file in three successive stages: 1) after the primary annihilation, 2) after all cascading ($n, p, \pi, \rho, \eta, \omega$) and evaporation particles have left the nucleus, and 3) after all decays of the meson resonances emitted from the nucleus ($\rho, \eta, \omega$) are modeled.

As discussed in the previous sections, multiplicity, charge, momentum and energy distributions of particles show good agreement with antiproton annihilation experimental data, and all simulated variables quantitatively satisfy the fundamental tenets of the required physics. Specifically, the generator has been shown to conserve charge, energy, momentum, baryon number, etc., through all three stages of simulation. The output file type is .txt, and formatted in such a way as to easily separate the particle content and their respective physical variables through the stages. Analysis of the output has been completed by ESS colleagues using C++ and the CERN ROOT 5.34 scientific software framework [54]. 100,000 simulated $\bar{n}C$ annihilation events are available upon request from the authors. The currently available event files and the following plots are created from the completed simulation file data without either antineutron or meson potentials.

An important characteristic for any relativistic many particle system is the invariant mass. One may analyze the invariant mass distribution for annihilation mesons at the annihilation point, and then see how it distorts due to interactions throughout the nucleus. Detector performance might affect the invariant mass further, but this study in not the focus of this paper. Fig. 11 shows how the distribution of invariant mass changes for all outgoing pions and photons generated by $\bar{n}C$ annihilation products (solid), a result of interactions with nuclear media. The dotted line shows the original distribution of invariant mass of the initial $\bar{n}N$ annihilation products. Fig. 11 shows that the intra-nuclear interactions of annihilation mesons with nucleons have resulted in a significant redistribution of energy between mesons and other nuclear constituents, shifting and smearing the initial distribution of $M_{inv}$ down to values of $\sim 1.2 \ GeV/c^2$. Note that the higher the initial value of $M_{inv}$, or the deeper the penetration of the antineutron into the nucleus, the larger the number of mesons which will interact with the nuclear environment, quickly devouring this particular part of the distribution.

Similarly, for Fig. 12, we see that the momentum distribution reconstructed from initial annihilation mesons is perturbed and expanded by transport through the nuclear environment. The structure shown in the dotted histogram shows a similar distribution as seen in Fig. 3, though implicitly convolved with Fig. 1, and considerate of different scales. After transport, this distribution distorts as particles cascade through the nucleus, shifting values up to as far as $\sim 0.8 \frac{GeV}{c}$.

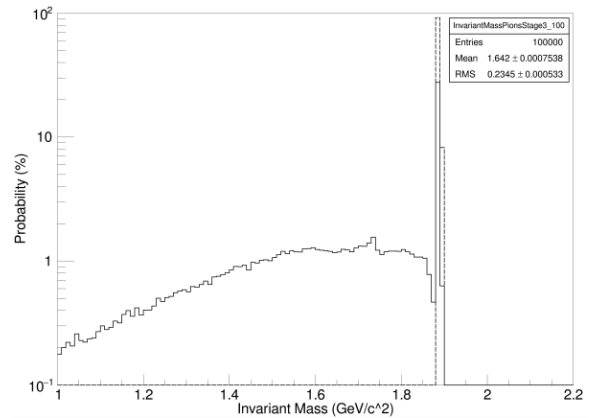

FIG 11. The distribution of invariant mass of $\bar{n}C$ annihilation products. The dotted histogram shows the distribution of invariant mass due only to original annihilation mesons. The solid histogram shows the invariant mass of pions and photons emanating from the nucleus after transport.

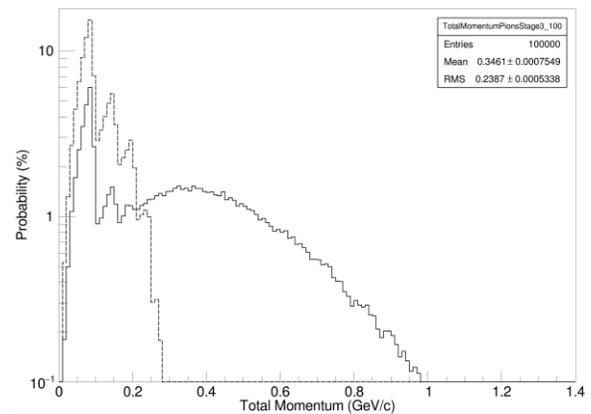

FIG. 12. The distribution of total momentum of $\bar{n}C$ annihilation products. The dotted histogram shows the distribution of total momentum of all original annihilation mesons. The solid histogram shows the distribution of total momentum of pions and photons emanating from the nucleus after transport.

Following discussion of Fig. 2 within [18], a highly relevant plot of total momentum versus invariant mass output variables is shown below in Fig. 13 for outgoing pions and photons. The projection of the $x$-axis is precisely Fig. 11, while the $y$-axis is Fig. 12. Across Figs. 11-13, all bin widths are identical (10 $MeV/c^2$ or $MeV/c$), and all counting scales are logarithmic. Note the bright spot at $\sim 1.9 \frac{GeV}{c^2}$ in invariant mass and $\sim 0.1 \frac{GeV}{c}$ in total momentum; this shape curves slightly upward and to the right, and contains the $\sim 35\%$ of all exiting pions and photons which go through the nucleus without interaction.

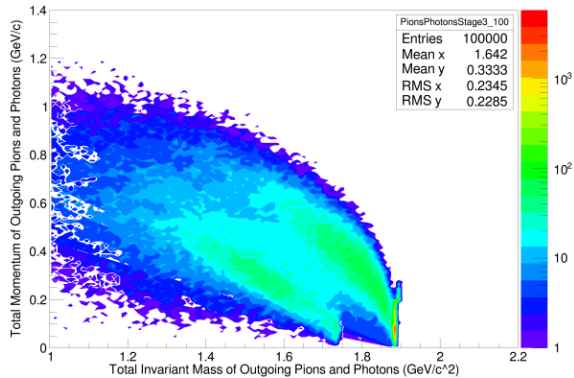

FIG. 13: Stage 3 total momentum vs. invariant mass of pions and photons. Note the double lobe structure; this is due to the absorption of a single pion during transport. Also recognize the thin, sickle-like shape in the lower right-hand corner of the figure; this is due to invariant mass of the initial Stage 1 mesons which did not interact with the nucleus.

## V.  CONCLUSION

This work has endeavored to demonstrate the detail of the optical-cascade model for describing antineutron annihilation on $^{12}_{6}C$ nuclei. It is quite important that the absorption of a slow antineutron is described within the framework of the optical model and that radial dependence of the annihilation probability is used within the initial stage of the simulation. A combination of experimental data with the results of a statistical model employing $SU(3)$ symmetry is used to describe the annihilation process. The propagation of annihilation-produced pions and heavier meson resonances within the nucleus is described by the intra-nuclear cascade model, which takes into account the nonlinear effect of decreasing nuclear density. The process of de-excitation of residual nuclei is described by a combination of the evaporation model and the Fermi model of explosive disintegration. This combined approach shows good agreement with experimental data in the modeling of antiproton annihilation at rest on $^{12}_{6}C$ nuclei, and provides a reliable predictor of the characteristics of slow antineutron annihilation on $^{12}_{6}C$. This model can thus be used as an event generator in the design of detector systems for planned experimental searches for neutron—antineutron transformation at the European Spallation Source, employing a free beam of cold neutrons.

This approach is universal, and can be used for simulating antineutron annihilation on many different nuclei. However, to search for the transformation occurring *within* nuclei (for example, within $^{40}_{18}Ar$, with no external source of antineutrons), a valid model can only be created when a proper definition of radial annihilation probability density is incorporated, allowing for the derivation of intra-nuclear $n \rightarrow \bar{n}$ transformation constraints. The model proposed in this work can be thought of as a first step in the preliminary modeling of this full process.

Expressing an aside into future developments, it is planned that more precise modeling will be rendered for the intra-nuclear $n \rightarrow \bar{n}$ transformation thought possibly to take place with $^{40}_{18}Ar$. Proper simulation of such a signal will be important to assess the feasibility of significantly improved transformation searches in the DUNE experiment. The goal of accurate and precise simulation is sought for the absolute suppression of the atmospheric neutrino background in the DUNE experiment, which will allow for an improvement in the search limits by several orders of magnitude. Inevitably, this will also require careful study and accurate simulation of atmospheric neutrino events in DUNE. The basic physics, model, file types, and analysis techniques presented in this article will continue to be employed, though special care must be taken in the integration of proper intra-nuclear transformation and radial annihilation probability distributions into the simulation. One other $n \rightarrow \bar{n}$ generator, already developed internally to the DUNE collaboration [23], is currently being studied by multiple colleagues in both the DUNE and ESS collaborations; complex techniques have been developed using neural networks and multivariate boosted decision trees to study the separability of supposed signal from atmospheric neutrino background with promising results. However, when considering the subtleties of the simulation assumptions and techniques, some room for improvements within the generator are thought to be possible. Thus, the independent generator development described in this work, along with its future $^{40}_{18}Ar$ extension, will help in understanding the potential and limits of exploration of rare processes like $n \rightarrow \bar{n}$ where separation from background plays a major role. Altogether, this will hopefully lead to a fruitful collaboration and collective meta-analysis between generators and groups, which could reputably assess the probability of separating signal from background sources in such a large, underground experiment.

### A. Acknowledgements

The authors would like to thank Jean-Marc Richard and Eduard Paryev for fruitful discussions of these nuclear processes, and to Yuri Kamyshkov for his permanent interest in this work. We wish to thank our colleagues in the NNBar ESS collaboration for the encouragement of this work. The work of JB was supported through DOE Grant DE-SC0014558 and the University of Tennessee Department of Physics. EG is grateful to the INR, Moscow for supporting her travel.